\documentclass[a4paper,aps,pop,twocolumn,superscriptaddress]{revtex4}

\usepackage{graphicx}
\usepackage{indentfirst}
\usepackage{units}
\usepackage{color}
\usepackage{hyperref}		
\usepackage[english]{babel}
\usepackage[utf8]{inputenc}
\usepackage{amsmath,amssymb,amsfonts}
\usepackage{float}

\begin{document}

\title{Ion firehose and ion cyclotron instability with subtracted-Kappa 
distributions
}

\author{Luiz F. Ziebell}
\email{luiz.ziebell@ufrgs.br}
\author{Rudi Gaelzer}
\email{rudi.gaelzer@ufrgs.br}
\affiliation{Instituto de F\'{\i}sica, Universidade Federal do Rio Grande do 
Sul, Caixa Postal 15051, 91501-970, Porto Alegre, RS, Brasil}
\author{Danny Summers}
\email{dsummers@mun.ca}
\affiliation{Memorial University of Newfoundland, St. John's, Newfoundland and
Labrador A1C 557, Canada}

\date{}

\begin{abstract}
In the present paper we discuss numerical solutions of the dispersion
relation for electromagnetic waves propagating along the lines of an
ambient magnetic field, considering parameters representative of space plasma
environments, and considering the ion population described by a 
subtracted-Kappa distribution. We consider situations in which the 
ion thermal anisotropy is such that $T_{i\perp}<T_{i\parallel}$, and 
investigate the effect of some parameters associated to the subtracted-Kappa
distribution on the ion firehose instability. We also consider situations
in which $T_{i\perp}>T_{i\parallel}$, and investigate the effects of the same
parameters on the ion-cyclotron instability. For both forms of instability,
we also investigate the effect of the thermal anisotropy of the electron
distribution, and the effect of the occurrence of a drift velocity in the
electron distribution function. 
Among the results obtained, we show that the
increase of the loss-cone feature in a bi-kappa distribution leads to a decrease
of the growth rates of the firehose instability, and to an increase of the
growth rates of the ion-cyclotron instability.
\end{abstract}


\maketitle

\section{Introduction}
\label{sec:introduction}

Plasmas are complex systems, in which waves and particles interact according
to rules relatively simple, but which may lead to a multitude of complex
phenomena. Space plasma environments, like the solar chomosphere, the solar
wind, or planetary magnetospheres, are examples of systems with sometimes
complex structures of ambient magnetic fields, presence of beams of particles
with a varied range of energies, occurrence of anisotropies in different plasma
parameters, and occurrence of inhomogeneities. In these environments, velocity
distributions are frequently far from the thermal equilibrium
\cite{PilippMMMRS87a,PilippMMMRS87b,MarschAT04,Marsch06,MatteiniLHPMVGM07,%
MaksimovicPL97,MaksimovicZCISLMMSLE05,PierrardML99,PierrardML01a,PierrardL10,%
StverakMTMFS09}. Due to their prevalence in space plasma environments, 
non thermal velocity distributions have been frequently employed in
investigations of waves and instabilites along the last decades
\cite{Vasyliunas68,Leubner04a,SummersT91,MaceHellberg95,Leubner02,Leubner04b,%
AliS21,ShaabanLW-SF21,MiceraBZSLLL20}.

Among the features frequently present in particle velocity distributions
in space environments, one of the most prominent is the presence of
energetic tails. These are frequently described by mathematical forms which
are known as Kappa distributions, which can be isotropic or anisotropic
\cite{Vasyliunas68,SummersT91,Leubner02,Leubner04b,bk:Livadiotis17,%
bk:LazarFichtner2021}. 
Anisotropic Kappa distribution can have isotropic kappa indexes and 
anisotropic thermal parameters, and are then known as bi-Kappa (BK) 
distributions, or can have also anisotropic kappa indexes, when they are
known as product-bi-Kappa distributions \cite{LazarP09a,LazarPS11b,%
LazarSP10,pl:SantosZG14}. Another feature which
may be found in magnetized space plasma environments is the depletion of
particles which constitutes the so-called loss-cone distributions. Other
ubiquitous feature in space environments is the presence of beams of
particles. All of these features may be significant for different forms of
wave instabilities. 

Summers and Stone \cite{SummersS2025} introduced a new velocity distribution
function in plasma physics called the substracted-Kappa (SK) distribution.
This distribution incorporates a loss-cone feature and an enhanced high-energy
tail characterized by the spectral index kappa. The SK distribution contains
two sources of free energy, namely the loss-cone feature and the thermal 
anisotropy, both of which can excite wave growth. In addition, Summers and
Stone \cite{SummersS2025} incorporated into the SK distribution a parallel
drift in the direction of the magnetic field. Because of its flexible
properties, as well as being set in a multi-dimensional parameter space, the
SK distribution has the potential for analyzing a wide variety of kinetic 
waves and microinstabilities in space physics. 

In the present paper, we utilize the SK distribution to further discuss
the dispersion relation for electromagnetic waves propagating parallel to
the ambient magnetic field. We utilize plasma parameters which are relevant
to plasma physics in space environments, consider the ion distribution 
described by a SK distribution, and discuss the effect of several
parameters, related to the SK distribution and to the velocity distribution
of the electron population. We consider situations in which the ion thermal
anisotropy is such that $T_{i\perp}/T_{i\parallel}<1$, which is the condition
conductive to the ion firehose instability, and also situations in which 
$T_{i\perp}/T_{i\parallel}>1$, which is the situation leading to the 
ion-cyclotron instability.
  
The plan of the paper is the following: In Sec. \ref{sec:theoretical} we 
present a very condensed discussion on the theoretical framework utilized
for the description of the system of interest for the investigation, and
obtain the dispersion relation. In Sec. \ref{sec:theoretical} we introduce
some basic parameters, and proceed with numerical solutions of the dispersion
relation. Results related to the firehose instability are presented and 
discussed in Sec. \ref{fh}, and results related to the ion-cyclotron 
instability are presented and discussed in Sec. \ref{ic}. Final comments are
presented at Sec. \ref{sec:final}.
 
\section{Theoretical Formulation}
\label{sec:theoretical}

In the present paper, the focus is on the study of electromagnetic waves
propagating along the
lines of an ambient magnetic field. The general dispersion relation for this 
kind of waves is obtained from the following determinant,
\begin{equation}
\label{determ}
\det \left( 
\begin{array}{ccc}
\epsilon_{xx}-N_{\parallel }^{2} & \epsilon_{xy} & 0 \\ 
\epsilon_{yx} & \epsilon_{yy}-N_{\parallel } & 0 \\ 
0 & 0 & \epsilon_{zz}
\end{array}
\right) =0~.  
\end{equation}
where the $\epsilon_{ij}$ are the components of the plasma dielectric tensor,
and $N_\parallel$ is the parallel component of the wave number vector,
${\bf N}= c{\bf k}/\omega$. It is seen that two different conditions can be
obtained, one associated to $\epsilon_{zz}=0$ (corresponding to electrostatic
waves) and the other associated to the upper left 2x2 determinant in Eq. 
(\ref{determ}), corresponding to electromagnetic waves, which are commonly 
denominated as Alfv\'en waves, in the low frequency regime.
This is the dispersion relation of interest for the present investigation.

Proceeding, we can introduce the plasma susceptibility
tensor $\chi_{ij}$, and write $\epsilon_{ij}=\delta_{ij}+\chi_{ij}$. Therefore, 
the dispersion relation for Alfv\'en waves becomes as follows, 
\begin{displaymath}
\left(1+\chi_{xx}^{C}-N_{\parallel }^{2}\right)
\left(1+\chi_{yy}^{C}-N_{\parallel }^{2}\right)
- \chi_{xy}^{C} \chi_{yx}^{C} = 0
\end{displaymath}

Using the symmetry conditions, $\chi_{yy}=\chi_{xx}$ and 
$\chi_{yx}=-\chi_{xy}$, and taking into account that $\chi_{xy}=i\xi_{xy}$, we
can write
\begin{displaymath}
\left(1+\chi_{xx}-N_{\parallel }^{2}\right)^2
- \xi_{xy}^2 = 0
\end{displaymath}
where the explicit expression for the susceptibility components are well known,
and can be written as follows,
\begin{eqnarray}
&&\chi_{xx} = \frac{1}{4 z^2} \sum_{a}
\frac{\omega_{pa}^2}{\Omega_\ast^2} \frac{1}{n_{a 0}} 
\sum_{n=-1,1} 
J(n,1,0;f_{a 0})\\
&&\chi_{xy} = i\,\frac{1}{4z^2} \sum_{a}
\frac{\omega_{pa}^2}{\Omega_\ast^2}\frac{1}{n_{a 0}}  
\sum_{n=-1,1} 
n J(n,1,0;f_{a 0})= i\,\xi_{xy},\nonumber
\end{eqnarray}
where the $J(n,m,h,f_{a0})$ are integral forms depending on the shape of the
equilibrium distribution function for particles of species $s$. These
$J$ integral forms are defined as follows \cite{pl:ZiebellSJG08,pl:ZiebellG17},
\begin{eqnarray}
\label{J,def}
&&J(n,m,h,f_{a0})= z \int d{\bf u}\frac{u_\parallel^hu_\perp^{2(m-1)}u_\perp
L(f_{a0})}{z-n\Omega_a/\Omega_*-q_\parallel u_\parallel}\\
&& L(f_{a,\kappa})= \frac{1}{\gamma}\left[
\left(\gamma-\frac{q_\parallel}{z}u_\parallel\right)
\frac{\partial}{\partial u_\perp}
+\frac{q_\parallel}{z}u_\perp \frac{\partial}{\partial u_\parallel}
\right]f_{a,\kappa},\nonumber
\end{eqnarray}
where $q_\parallel$ is the normalized wavenumber, 
$q_\parallel= k_\parallel v_*/\Omega_*$, and $z$ is the normalized
angular wave frequency, $z=\omega/\Omega_*$. 
The quantitites $v_*$ and $\Omega_*$ are a
characteristic speed and a characteristic frequency, respectively, which
can be chosen for convenience of the description of the problem investigated.

Proceeding, the dispersion relation leads to 
$1+\chi_{xx}-N_{\parallel }^{2}=\pm\xi_{xy}$, and 
therefore
\begin{eqnarray*}
&&N_{\parallel }^{2} = 1+\frac{1}{4 z^2} \sum_{a}
\frac{\omega_{pa}^2}{\Omega_\ast^2} \frac{1}{n_{a 0}}\\
&&\times 
\left[
J(1,1,0;f_{a 0})
+J(-1,1,0;f_{a 0})\right.\\
&&\left.
\mp\left(
J(1,1,0;f_{a 0})
-J(-1,1,0;f_{a 0})
\right)
\right]
\end{eqnarray*}

It is seen that the dispersion relation can be written in two different forms,
according to the sign $\pm$ which is utilized. Introducing sign $s=\pm 1$,
and using the dimensionless variables, one obtains
\begin{equation}
\label{disprel}
\frac{c^2}{v_*^2} q_{\parallel}^{2}
= z^2+\frac{1}{2} \sum_{a}
\frac{\omega_{pa}^2}{\Omega_\ast^2} \frac{1}{n_{a 0}} 
J(s,1,0;f_{a 0}).
\end{equation}

To evaluate the dispersion relation we choose the SK distribution incorporating
a parallel drift orginally developed by Summers and Stone \cite{SummersS2025},
namely,
\begin{eqnarray}
\label{fa,BKd}
&&f_{a}^{(SK)}\left(u_\parallel,u_\perp; u_{a\kappa\parallel},
u_{a\kappa\perp}\right)=\\
&&\times\frac{n_{a 0}}{\pi^{3/2}\kappa^{3/2}
u_{a\kappa\parallel}u_{a\kappa\perp}^2}
\frac{\Gamma(\kappa_a+1)}{\Gamma(\kappa_a-1/2)}\nonumber\\
&&\times
\left[\left(\frac{1-\beta_a\Delta_a}{1-\beta_a}\right)
\left(1+\frac{(u_{\parallel}-u_{da})^{2}}{\kappa_{a}u_{a\kappa\parallel}^{2}}
+\frac{u_{\perp}^{2}}{\kappa_{a}u_{a\kappa\perp}^{2}}
\right)^{-(\kappa_{a}+1)}\right.\nonumber\\
&&\left. -\left(\frac{1-\Delta_a}{1-\beta_a}\right)
\left(1+\frac{(u_{\parallel}-u_{da})^{2}}{\kappa_{a}u_{a\kappa\parallel}^{2}}
+\frac{u_{\perp}^{2}}{\beta_a\kappa_{a}u_{a\kappa\perp}^{2}}
\right)^{-(\kappa_{a}+1)}\right],\nonumber
\end{eqnarray}
where $u_\parallel$ and $u_\perp$ are the parallel and perpendicular
components of the normalized velocity normalized, ${\bf u}=
{\bf v}/v_*$, and where $u_{da}$ is a normalized drift velocity. 

For computational and algebraic purposes, it may be useful to express
distribution (\ref{fa,BKd}) as follows:
\begin{eqnarray}
\label{fa,SKd}
&&f_{a}^{(SK)}\left(u_\parallel,u_\perp\right)=
\frac{1-\beta_a\Delta_a}{1-\beta_a} 
f_{a}^{(BK)}\left(u_\parallel,u_\perp,u_{a\kappa\parallel}.
u_{a\kappa\perp}\right)\nonumber\\
&&-\frac{1-\Delta_a}{1-\beta_a}\beta_a 
f_{a}^{(BK)}\left(u_\parallel,u_\perp,u_{a\kappa\parallel},
\sqrt{\beta_a}u_{a\kappa\perp}\right),
\end{eqnarray}
where
\begin{eqnarray}
\label{fa,BK}
&&f_{a}^{(BK)}\left(u_\parallel,u_\perp; u_{a\kappa\parallel},
u_{a\kappa\perp}\right)=\\
&&\times\frac{n_{a 0}}{\pi^{3/2}\kappa^{3/2}
u_{a\kappa\parallel}u_{a\kappa\perp}^2}
\frac{\Gamma(\kappa_a+\alpha_a)}{\Gamma(\kappa_a+\alpha_a-3/2)}\nonumber\\
&&\times
\left(1+\frac{(u_{\parallel}-u_{da})^{2}}{\kappa_{a}u_{a\kappa\parallel}^{2}}
+\frac{u_{\perp}^{2}}{\kappa_{a}u_{a\kappa\perp}^{2}}
\right)^{-(\kappa_{a}+\alpha_a)}\nonumber
\end{eqnarray}
is a bi-Kappa distribution. The parameter $\alpha_a$, here interpreted as a
free numerical parameter, was taken as 1 in the original definition 
\eqref{fa,BKd} of the SK distribution. In the general historical context of
Kappa distributions, $\alpha_a$ has been tipically set as 0 or 1
(Refs. \cite{Vasyliunas68,SummersT91,MaceHellberg95,Leubner02,Leubner04b,%
LazarFY16}. 
If the parameter $\alpha_a$ is taken as
$\alpha_a=1$, then $\displaystyle u_{a\kappa\parallel}^2
=\frac{\kappa_a-3/2}{\kappa_a} u_{a\parallel}^2$ and
$\displaystyle u_{a\kappa\perp}^2
=\frac{\kappa_a-3/2}{\kappa_a}u_{a\perp}^2$, where $u_{a\parallel}^2$ and
$u_{a\perp}^2$ are $\kappa$-independent constants. If the parameter
$\alpha_a$ is taken as $\alpha_a=0$, then
$\displaystyle u_{a\kappa\parallel}^2=u_{a\parallel}^2$,
$\displaystyle u_{a\kappa\perp}^2=u_{a\perp}^2$.

Normalization of both the SK and BK distributions is carried out by
specifying $\int f_a d^3v=n_{a0}$.

Equation (\ref{fa,SKd}) is a very flexible form of distribution, which allows
for descriptions of a wide variety of different situations of interest in 
physics of plasmas \cite{SummersS2025}. It becomes a BK distribution for 
$\Delta_a=1.0$, for any value of $\beta_a$. It also becomes a BK distribution
in the case of $\beta_a=0.0$, for any value of $\Delta_a$. For $\Delta_a=0.0$ and
$\beta_a\to 1.0$, the SK tends to a drifting bi-Kappa loss-cone distribution.
That is, for $\Delta_a=0.0$, the SK changes from a drifting bi-Kappa 
distribution to a drifting bi-Kappa loss cone as $\beta_a$ changes between 0.0
and 1.0. Different degrees of loss-cone filling can be obtained in the
range $0\le \Delta_a < 1.0$, for $1.0< \beta_a< 1.0$. It is therefore seen
that the SK is a very suitable form of distribution for the investigation
of wave phenomena in different plasma physics environments. 

According to the objective of the present paper, we have to solve the
dispersion relation (\ref{disprel})
for cases in which at least one of the plasma species
is described by a SK distribution, i.e., $f_{a0}=f_a^{(SK)}$. 
It is then necessary to evaluate the 
required quantity, $J(s,1,0,f_{a}^{(SK)})$,
\begin{eqnarray}
\label{J,h=0}
&&J(s,1,0;f_a^{(SK)})\\
&&=\frac{1-\beta_a\Delta_a}{1-\beta_a} J(s,1,0;f_a^{(BK)}(u_\parallel,u_\perp;
u_{a\kappa\parallel},u_{a\kappa\perp}))\nonumber\\
&&- \frac{1-\Delta_a}{1-\beta_a}\beta_a J(s,1,0;f_a^{(BK)}(u_\parallel,u_\perp;
u_{a\kappa\parallel},\sqrt{\beta_a}u_{a\kappa\perp})).\nonumber
\end{eqnarray}
Details of the calculation can be found in the Appendix.

\section{Numerical Analysis}
\label{sec:numerical}

\subsection{Firehose instability}
\label{fh}

For the numerical analysis, we consider a plasma with two particle populations,
ions and electrons, and consider that the ions are protons, so that $m_i=m_p$.
As basic parameters, we consider $v_{i\parallel}^2/v_A^2=2.0$ and
$v_A/c= 1.0\times 10^{-4}$, with $v_{i\parallel}$ being the ion parallel
thermal velocity, and $v_A=B_0/\sqrt{4\pi n_{i0}m_i}$ being the Alfv\'en 
velocity. These values can be considered as representative of space plasma
environments \cite{Gary05}.
We also consider for all cases $T_{e\parallel}=T_{i\parallel}$.
Moreover, we assume $v_*=v_A$ and $\Omega^*=\Omega_i$, so that 
$z=\omega/\Omega_i$, and $q_\parallel=k_\parallel v_A/\Omega_i$.
We then solve the dispersion relation (\ref{disprel}) considering 
different forms for the distribution functions $f_{a0}$, and different values
for some plasma parameters. We start with the investigation about the firehose
instability, which is characterized by $T_{i\perp}/T_{i\parallel}<1.0$.

\begin{widetext}

\begin{figure}[H]
\includegraphics{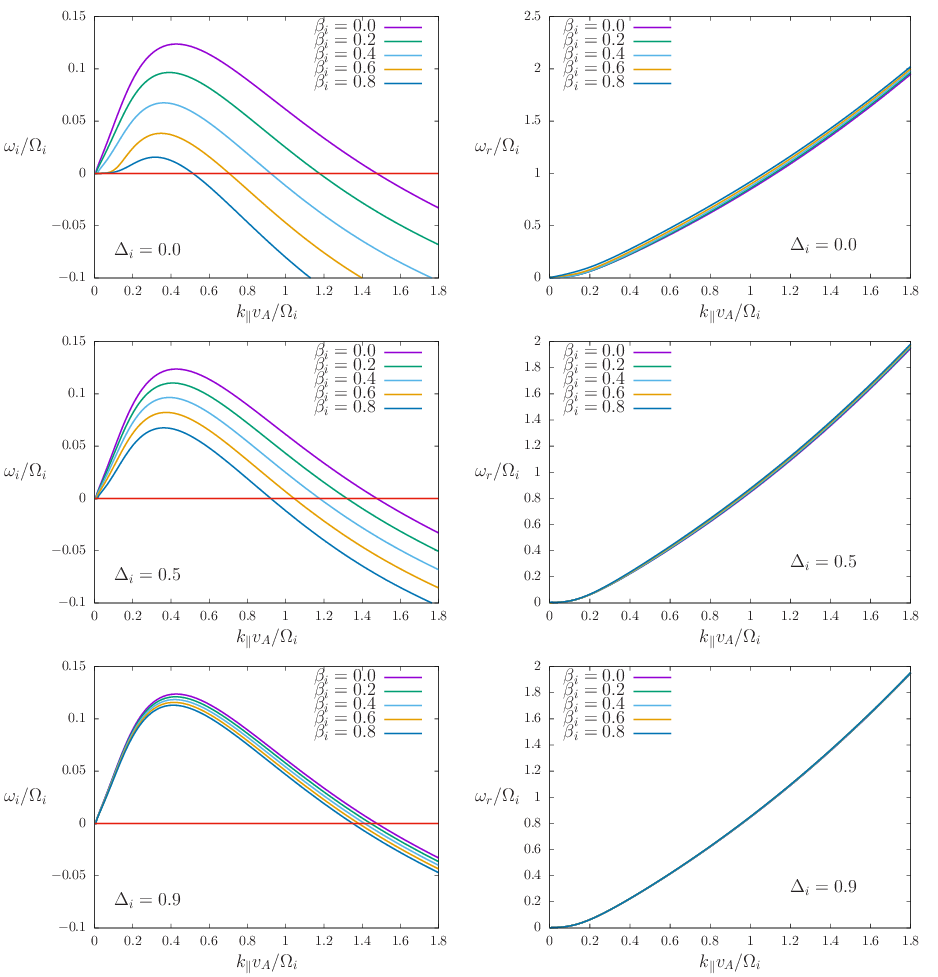}
\caption{
The left panels show the imaginary part of the normalized frequency of unstable
parallel propagating electromagnetic waves, $z_i$, vs. normalized wave number 
$q_\parallel$. 
The panels to the right show the corresponding real part of the normalized 
frequency, $z_r$, vs. normalized wave number $q_\parallel$.
Ions with a subtracted-Kappa distribution (SK), with $\kappa_i=3.0$, 
and $T_{i\perp}/T_{i\parallel}= 0.4$.
Electrons with a BK distribution, with $\kappa_e=3.0$ and
$T_{e\perp}/T_{e\parallel}= 0.4$.
Top panels: $\Delta_i=0.0$, and several values of $\beta_i$;
middle panels: $\Delta_i=0.5$, and several values of $\beta_i$;
bottom panels $\Delta_i=0.9$, and several values of $\beta_i$.
}
\label{fig1}
\end{figure}
\end{widetext}

In Figure \ref{fig1} we show the imaginary and real parts of the normalized 
frequency $z$ obtained from numerical solution of the dispersion relation, for
parallel propagating electromagnetic waves which show unstable behavior, i.e.
$z_i>0$, in the range of wave numbers which is investigated. The left panels
of Fig. \ref{fig1} show the values of $z_i$, and the right panels show the 
values of $z_r$, versus normalized wave number $q_\parallel$. 
For this analysis, the ions are assumed with a subtracted-Kappa 
distribution (SK), with $\kappa_i=3.0$, and $T_{i\perp}/T_{i\parallel}= 0.4$.
Electrons are assumed to have a BK distribution, with $\kappa_e=3.0$ and
$T_{e\perp}/T_{e\parallel}= 0.4$. 
The panels at the top of Fig. \ref{fig1} show curves obtained in the case
$\Delta_{i}=0.0$, for several values of $\beta_{i}$. It is seen that
the range of unstable wave number values decrease by a factor $\simeq 3$, 
and the magnitude of $z_i$ (the growth rate) decreases by almost one order
of magnitude, while $\beta_i$ changes from 0.0 to 0.8 (i.e., while the
ion distribution changes from a BK to a SK). On the other hand, the right 
panel shows that the value of $z_r$ is little affected by the change of
$\beta_i$. The middle panels show results obtained in the case of 
$\Delta_i=0.5$, and several values of $\beta_i$. The middle left
panel shows that the values of $z_i$ are reduced by a factor of $\simeq 2$
between $\beta_i=0.0$ and 0.8, while the range of unstable wave numbers
is reduced by a factor smaller than 2. The middle right panel shows that
the values of $z_r$ are even less affected by the change of $\beta_i$
than in the case of $\beta_i=0.0$. 
Finally the bottom panels show that both $z_i$ and $z_r$ are little 
affected by the change of $\beta_i$. This is hardly surprising, since 
in the limit $\Delta_i=1.0$ the ion distribution becomes independent
of $\beta_i$. 

\begin{widetext}

\begin{figure}[H]
\includegraphics{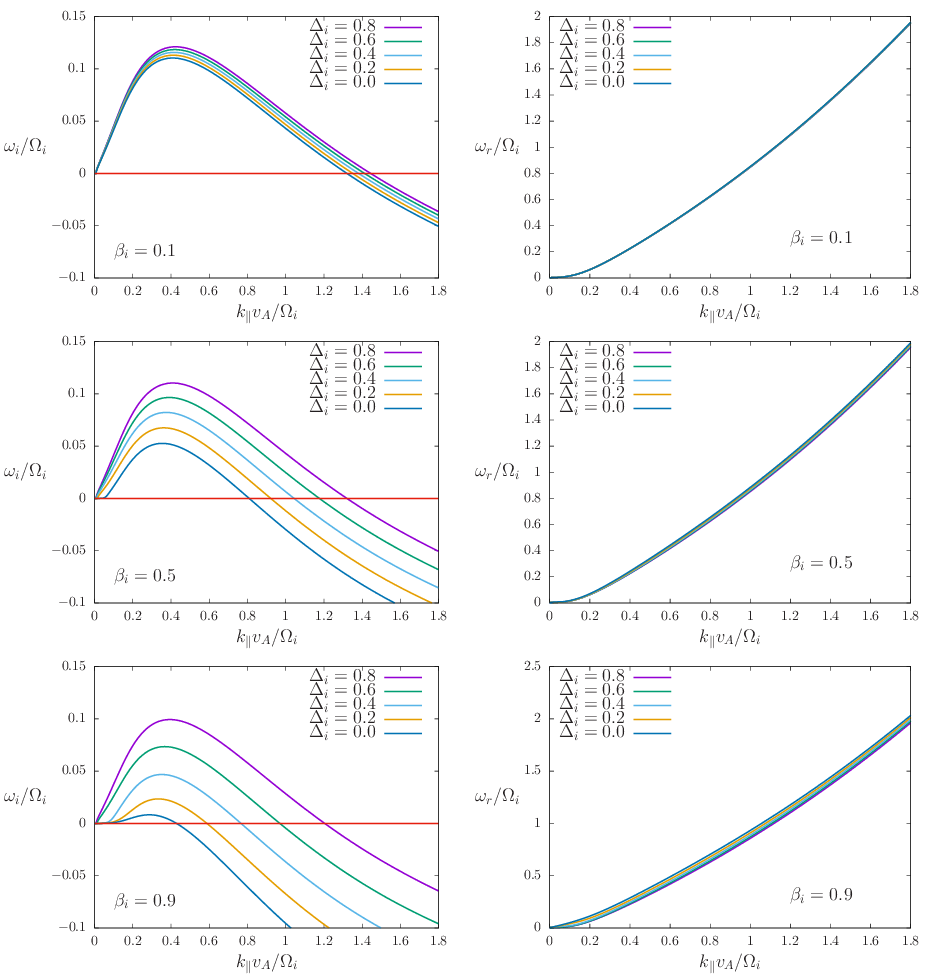}
\caption{
The left panels show the imaginary part of the normalized frequency of unstable
parallel propagating electromagnetic waves, $z_i$, vs. normalized wave number 
$q_\parallel$. 
The panels to the right show the corresponding real part of the normalized 
frequency, $z_r$, vs. normalized wave number $q_\parallel$.
Ions with a subtracted-Kappa distribution (SK), with $\kappa_i=3.0$, 
and $T_{i\perp}/T_{i\parallel}= 0.4$.
Electrons with a BK distribution, with $\kappa_e=3.0$ and
$T_{e\perp}/T_{e\parallel}= 0.4$.
Top panels: $\beta_i=0.1$, and several values of $\Delta_i$;
middle panels: $\beta_i=0.5$, and several values of $\Delta_i$;
bottom panels: $\beta_i=0.9$, and several values of $\Delta_i$.
}
\label{fig2}
\end{figure}
\end{widetext}

Figure \ref{fig2} shows the values of $z_i$ and $z_r$ for the same 
distributions and parameters considered in the case of Fig. \ref{fig1},
but considering a range of values of $\Delta_i$, for fixed values of
$\beta_i$.
As in Fig. \ref{fig1}, values of $z_i$ appear in the left panels
of Fig. \ref{fig2}, versus normalized wave number $q_\parallel$,
and the values of $z_r$ are in the right panels.
The ions are assumed with a subtracted-Kappa 
distribution (SK), with $\kappa_i=3.0$ $T_{i\perp}/T_{i\parallel}= 0.4$,
and electrons have a BK distribution, with $\kappa_e=3.0$ and
$T_{e\perp}/T_{e\parallel}= 0.4$. 
The top panels of Fig. \ref{fig2} show the case of $\beta_{i}=0.1$, for 
several values of $\Delta_{i}$, between 0.0 and 0.8. It is seen that
the values of $z_i$ are little affected by the change of $\Delta_i$, 
with a small reduction in magnitude and in the extension ot the region of
unstable wave numbers, while $\Delta_i$ is decreased from 0.8 to 0.0.
The effect is more noticeable for larger values of $\beta_i$. The panels
at the middle row of Fig. \ref{fig2} show that the decrease of $\Delta_i$
leads to decrease of the instability, both in magnitude of the growth rate
and in the range of unstable wave numbers. The effect is such that the
maximum growth rate obtained with $\Delta_i=0.0$ is nearly one-half of the
maximum growth rate obtained with $\Delta_i=0.8$.
For larger values of $\beta_i$, the effect of $\Delta_i$ becomes even more
significant. This is illustrated by the case of $\beta_i=0.9$, at the
bottom line of Fig. \ref{fig2}, which shows that the instability almost
vanishes for $\Delta_i\to 0$. Regarding the values of $z_r$, the top panels
show that they are nearly insensitive to $\Delta_i$, in the case of small
$\beta_i$. For increasing values of $\beta_i$, the effect of $\Delta_i$ is
small, but becomes more significant, with the general tendency of increasing 
in the value of the frequency at each wave number, with the increase of 
$\Delta_i$.

\begin{widetext}

\begin{figure}[H]
\includegraphics{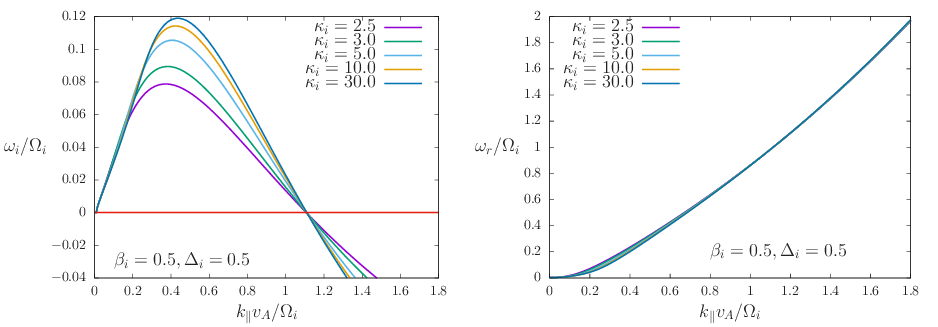}
\caption{
The left panel shows the imaginary part of the normalized frequency of unstable
parallel propagating electromagnetic waves, $z_i$, vs. normalized wave number 
$q_\parallel$. The 
panels to the right show the corresponding real part of the normalized 
frequency, $z_r$, vs. normalized wave number $q_\parallel$.
Ions with a subtracted-Kappa distribution (SK), with 
$T_{i\perp}/T_{i\parallel}= 0.4$, $\beta_i=0.5$, and 
$\Delta_i=0.5$.
Electrons with a BK distribution, with $\kappa_e=3.0$ and
$T_{e\perp}/T_{e\parallel}= 0.4$.
The panels show curves corresponding to several values of $\kappa_i$ ($\kappa_i
$=2.5, 2.0, 5.0, 10.0, and 30.0).
}
\label{fig3}
\end{figure}
\end{widetext}

In the sequence, we investigate the effect of the parameter $\kappa_i$ on the
firehose instability. Figure 3 shows the values of $z_i$ and $z_r$ for unstable
parallel propagating electromagnetic waves, considering a situation where
the ion population is described by a subtracted-Kappa distribution (SK), 
with $T_{i\perp}/T_{i\parallel}= 0.4$, $\beta_i=0.5$, and $\Delta_i=0.5$, and
the electron population is described by a BK distribution, with $\kappa_e=3.0$ 
and $T_{e\perp}/T_{e\parallel}= 0.4$. With the chosen values for $\beta_i$ and
for $\Delta_i$, the ion distribution exhibits a partially filled loss-cone.
With this basic configuration, we 
consider five different values of the ion kappa parameter, by taking
$\kappa_i=2.5$, 2.0, 5.0, 10.0, and 30.0. The right panel of Fig. \ref{fig3}
shows that the real part of the wave frequency is little affected by the change
of $\kappa_i$. That is, $z_r$ is little affected by the nature of the 
extended tail of the ion distribution function. On the other hand, the
imaginary part of the wave frequency shows significant effect of the $\kappa_i$
parameter. The panel at the left of Fig. \ref{fig3} shows that the maximum
value of $z_i$ changes from $z_i\simeq 0.075$ in the case of $\kappa_i=2.5$ 
to $z_i\simeq 0.12$ for $\kappa_i=30.0$, situation in which the distribution
is close to the subtracted-Maxwellian limit.

\begin{widetext}

\begin{figure}[H]
\includegraphics{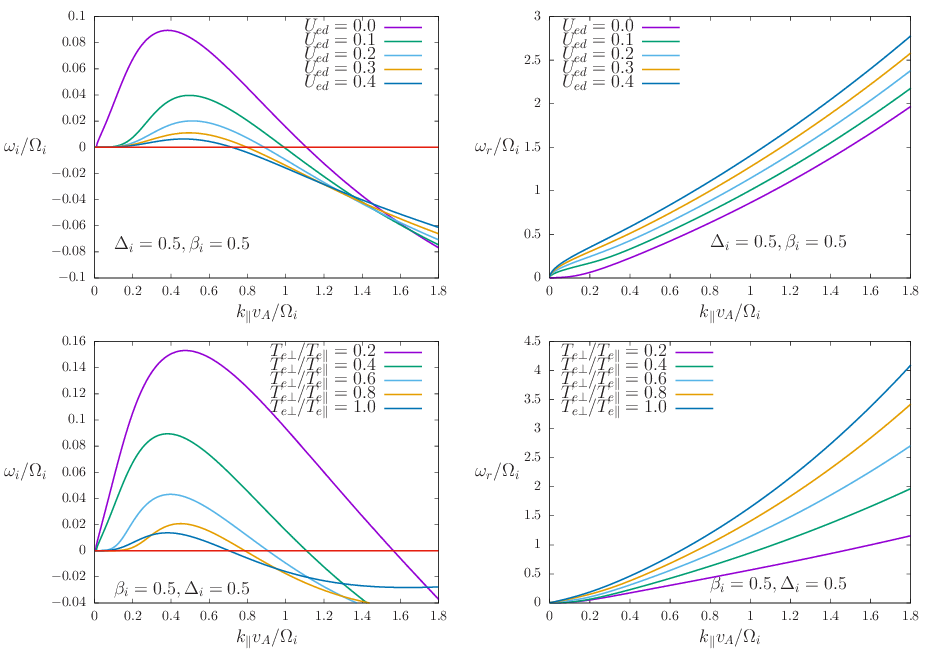}
\caption{
The left panels show the imaginary part of the normalized frequency of unstable
parallel propagating electromagnetic waves, $z_i$, vs. normalized wave number 
$q_\parallel$. 
The panels to the right show the corresponding real part of the normalized 
frequency, $z_r$, vs. normalized wave number $q_\parallel$.
Ions with a subtracted-Kappa distribution (sK), with $\kappa_i=3.0$, 
$T_{i\perp}/T_{i\parallel}= 0.4$, $\beta_i=0.5$, and $\Delta_i=0.5$;
Top panels: Electrons with a drifting BK distribution, with $\kappa_e=3.0$ and
$T_{e\perp}/T_{e\parallel}= 0.4$, for several values of the normalized electron 
drift velocity, $U_{ed}$. 
Bottom panels: Electrons with a BK distribution ($U_{ed}=0$), with 
$\kappa_e=3.0$, and several values of the temperature ratio
$T_{e\perp}/T_{e\parallel}$.
}
\label{fig4}
\end{figure}
\end{widetext}

Before closing the discussion on the ion-firehose instability, we investigate
the effect of characteristics of the electron distribution function on the
instability, which is caused by anisotropy of the ion distribution.
The first analysis is made considering the presence of a drifting velocity
in the electron population. The top panels of Fig. \ref{fig4} show results
obtained with ions featuring a subtracted-Kappa distribution (SK), with 
$\kappa_i=3.0$, $T_{i\perp}/T_{i\parallel}= 0.4$, $\beta_i=0.5$, and 
$\Delta_i=0.5$, and electrons described by a drifting BK distribution,
with $\kappa_e=3.0$ and the same temperature anisotropy as in the ion
distribution, $T_{e\perp}/T_{e\parallel}= 0.4$, considering several values of
the normalized electron drift velocity, $U_{ed}$. The effect of the electron
drift velocity is to reduce the magnitude of the instability, so that the 
magnitude of the highest growth rate is reduced by nearly a factor 2 between
$U_{ed}=0.0$ and $U_{ed}=0.1$, and by nearly a factor 14 between 0.0 and 0.4.
On the other hand, the value of $z_r$ is significantly affected by the
change in the electron drifting velocity, at all wavelengths of the unstable
region, as illustrated by the top right panel of Fig. \ref{fig4}. 

In the bottom panels of Fig. \ref{fig4}, we show results obtained considering
the same SK ion distribution utilized for the top panels, with a BK 
distribution without drift velocity for the electrons, with different values
of the electron anisotropy parameter $T_{e\perp}/T_{e\parallel}$. The different
curves shown in the bottom panels therefore consider cases with a fixed 
anisotropy in the ion distribution, and changing anisotropy in the electron
distribution. The left panel shows that in the case of isotropic electron
distribution the maximum growth rate is $z_i\simeq 0.013$, for 
$T_{e\perp}/T_{e\parallel}=0.4$ (i.e., the same anisotropy ratio as in the ion
distribution) the maximum value is $z_i\simeq 0.089$, and for 
$T_{e\perp}/T_{e\parallel}=0.2$ the maximum growth rate is $i_i\simeq 0.15$.
The range of unstable wave numbers is also seen to increase with the increase
of $T_{e\perp}/T_{e\parallel}$.
That is, the increase in the anisotropy in the electron distribution 
contributes to the increase of the ion-firehose instability. This conclusion
obtained here considering ions with a SK distribution is similar to what
has been observed in the case of ions with bi-Kappa distributions, without the
loss-cone feature \cite{pl:ZiebellG19}. The real part of the wave frequency
is also affected by the change of the electron anisotropy. The right bottom
panel of Fig. \ref{fig4} shows that the value of $z_r$ decreases with the 
increase of $T_{e\perp}/T_{e\parallel}$, for all wavelenghts in the range which
has been investigated.

\subsection{Ion-cyclotron instability}
\label{ic}

For the investigation on the ion-cyclotron instability, we consider the same
dispersion relation and basic parameters utilized in the case of the 
firehose instability, in
the previous sub-section, but assume $T_{i\perp}/T_{i\parallel}>1.0$. 

\begin{widetext}

\begin{figure}[H]
\includegraphics{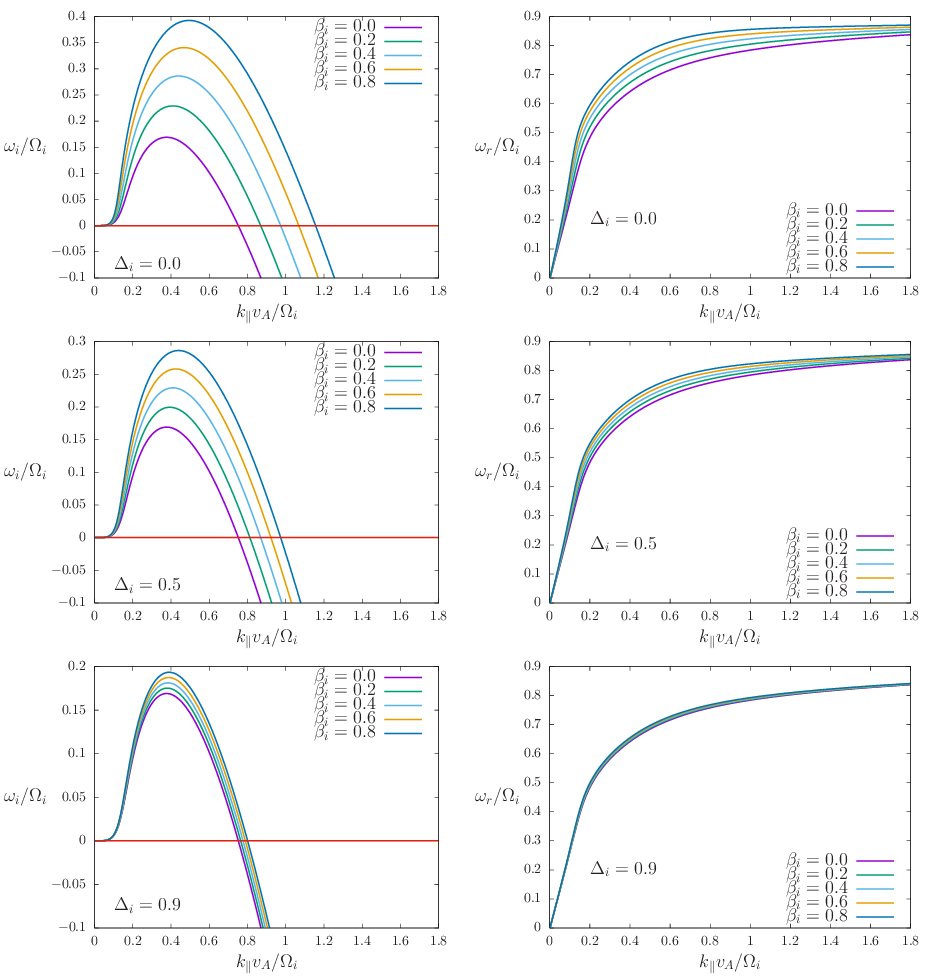}
\caption{
The left panels show the imaginary part of the normalized frequency of unstable
parallel propagating electromagnetic waves, $z_i$, vs. normalized wave number 
$q_\parallel$. 
The panels to the right show the corresponding real part of the normalized 
frequency, $z_r$, vs. normalized wave number $q_\parallel$.
Ions with a subtracted-Kappa distribution (SK), with $\kappa_i=3.0$, 
and $T_{i\perp}/T_{i\parallel}= 4.0$.
Electrons with a BK distribution, with $\kappa_e=3.0$ and
$T_{e\perp}/T_{e\parallel}= 4.0$.
Top: $\Delta_i=0.0$, and several values of $\beta_i$;
middle: $\Delta_i=0.5$, and several values of $\beta_i$;
bottom: $\Delta_i=0.9$, and several values of $\beta_i$.
}
\label{fig5}
\end{figure}
\end{widetext}

Figure \ref{fig5} displays the values of $z_i$ and $z_r$ for the unstable waves
obtained from numerical solution of the dispersion relation. 
As in the case of the analysis on the firehose instability, the ions are
assumed with a subtracted-Kappa 
distribution (SK), with $\kappa_i=3.0$, but for the ion-cyclotron
instability investigation we assume $T_{i\perp}/T_{i\parallel}= 4.0$ as a 
representative value.
Electrons are assumed to have a BK distribution, with $\kappa_e=3.0$ and
$T_{e\perp}/T_{e\parallel}= 4.0$. 
The top panels of Fig. \ref{fig5} show the case
$\Delta_{i}=0.0$, for several values of $\beta_{i}$. It is seen that the
upper limit of the range of unstable wave number moves from $q\simeq 0.75$ to
$q\simeq 1.18$, as $\beta_i$ changes from 0.0 to 0.8, and the maximum value
of $z_i$ changes from 0.17 to 0.38. That is, the increase of $\beta_i$ (which
means the change from BK to SK in the ion distribution)
contributes significantly to the increase of the ion-cyclotron instability.
The real part of the wave frequency is also affected by the change of 
$\beta_i$, as shown in the top right panel, where the $z_r$ are seen to
increase with the increase of $\beta_i$.
These effects of the effect of $\beta_i$, for fixed value of $\Delta_i$,
are qualitatively similar to those exhibited by the left panel in Fig. 5
of Ref. \cite{SummersS2025}, where some results obtained from numerical
solution of the dispersion relation of parallel propagating electromagnetic
waves were presented, for different parameters.

Effects which are qualitatively similar, but less intense, are seen in the 
case of larger values of $\Delta_i$ (i.e., for increasingly filled loss-cone
in the ion distribution), as seen in the middle and bottom panels of Fig.
\ref{fig5}. In fact, the bottom panels show that in the case of 
$\Delta_i=0.9$ the quantity $z_r$ is practically unaffected by the change
of $\beta_i$, and the corresponding value of $z_i$ is increased from 
$z_i\simeq 0.17$ for $\beta_i=0.0$ to $z_i\simeq 0.19$ for $\beta_i=0.8$.

\begin{widetext}

\begin{figure}[H]
\includegraphics{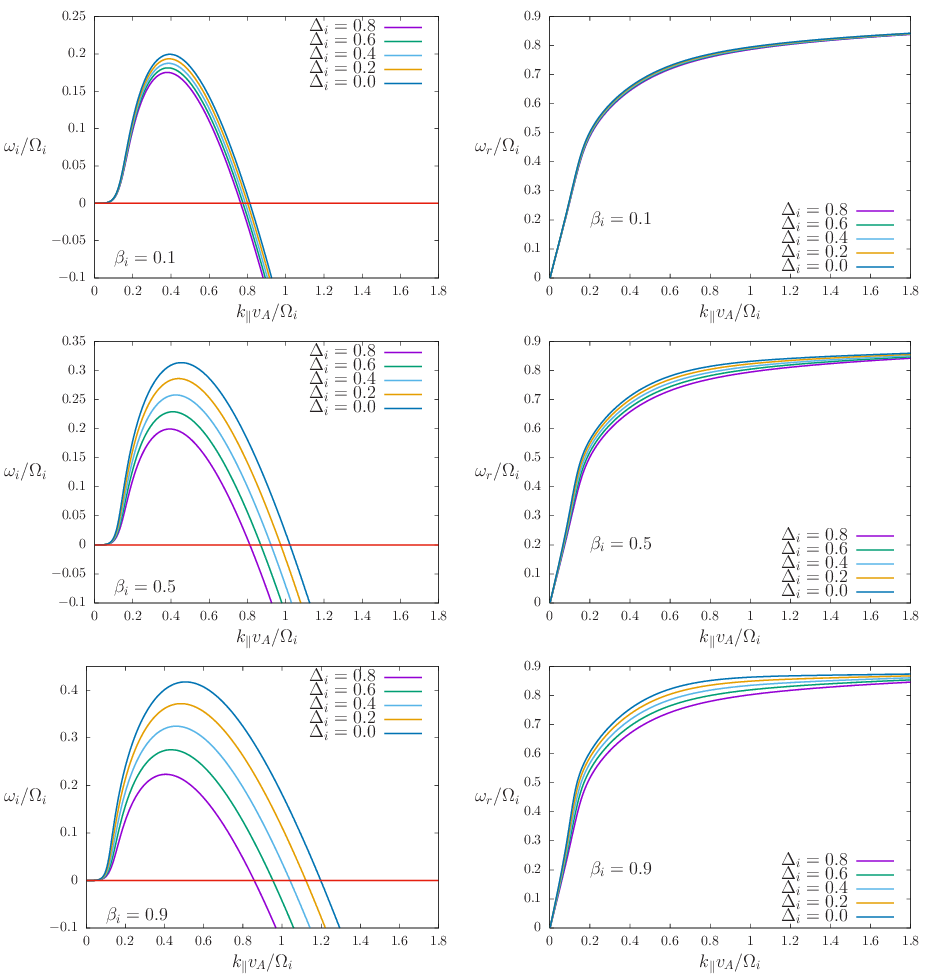}
\caption{
The left panels show the imaginary part of the normalized frequency of unstable
parallel propagating electromagnetic waves, $z_i$, vs. normalized wave number 
$q_\parallel$. 
The panels to the right show the corresponding real part of the normalized 
frequency, $z_r$, vs. normalized wave number $q_\parallel$.
Ions with a subtracted-Kappa distribution (SK), with $\kappa_i=3.0$, 
and $T_{i\perp}/T_{i\parallel}= 4.0$.
Electrons with a BK distribution, with $\kappa_e=3.0$ and
$T_{e\perp}/T_{e\parallel}= 4.0$.
Top: $\beta_i=0.1$, and several values of $\Delta_i$;
middle: $\beta_i=0.5$, and several values of $\Delta_i$.
bottom: $\beta_i=0.9$, and several values of $\Delta_i$.
}
\label{fig6}
\end{figure}
\end{widetext}

Figure \ref{fig6} shows results obtained with fixed values of $\beta_i$, and
several values of $\Delta_i$. In the top panels, the results are obtained
considering $\beta_i=0.1$. The right panel shows that the value of $z_r$ is
practically unaffected in this case, and the left panel shows that $z_i$ is
slightly decreased with the increase of $\Delta_i$ (which means the filling-up
of the loss-cone). For increasing values of $\beta_i$ the effect of 
$\Delta_i$ becomes more pronounced, both for the real part $z_r$ and for the
imaginary part $z_i$. For instance, the bottom left panel of Fig.
\ref{fig6} shows that in the case of $\beta_i=0.9$ the maximum growth rate 
is decreased by nearly a factor of 2 when $\Delta_i$ is changed from 0.0
to 0.8, and the bottom right panel shows that the frequency of the maximum
instability, at $q\simeq 0.5$, decreases in such a way that $z_r\simeq 0.75$
for $\Delta_i=0.0$ and is $z_r\simeq 0.65$ for $\Delta_i=0.8$. 
These results are qualitatively similar to those presented in the middle panel 
of Fig. 5 in Ref. \cite{SummersS2025}, obtained considering different 
parameters.

\begin{widetext}

\begin{figure}[H]
\includegraphics{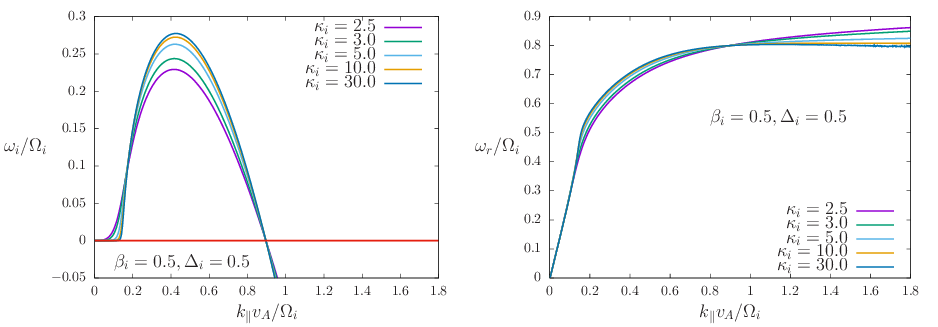}
\caption{
The left panel shows the imaginary part of the normalized frequency of unstable
parallel propagating electromagnetic waves, $z_i$, vs. normalized wave number 
$q_\parallel$. 
The panels to the right show the corresponding real part of the normalized 
frequency, $z_r$, vs. normalized wave number $q_\parallel$.
Ions with a subtracted-Kappa distribution (SK), with 
$T_{i\perp}/T_{i\parallel}= 4.0$, $\beta_i=0.5$, and 
$\Delta_i=0.5$.
Electrons with a BK distribution, with $\kappa_e=3.0$ and
$T_{e\perp}/T_{e\parallel}= 4.0$.
The panels show curves corresponding to several values of $\kappa_i$ ($\kappa_i
$=2.5, 2.0, 5.0, 10.0, and 30.0).
}
\label{fig7}
\end{figure}
\end{widetext}

In Fig. \ref{fig7} we investigate the effect of the $\kappa_i$ parameter on 
the growth-rate and wave frequency of the ion-cyclotron instability.
As in the case of Fig. \ref{fig3} for the firehose instability, we consider
a fixed electron BK distribution, without drift velocity, with $\kappa_e=3.0$,
and assume $T_{e\perp}/T_{e\parallel}= 4.0$. 
The ion distribution is supposed to be a subtracted-Kappa distribution 
(SK), with $T_{i\perp}/T_{i\parallel}= 0.4$, $\beta_i=0.5$, and $\Delta_i=0.5$,
and the $\kappa_i$ parameter is changed between 2.5 and 30.
Figure \ref{fig7} shows in the right panel that the value of $z_r$ increases
with the increase of the Maxwellian character of the ion distribution, that is,
with the increase of $\kappa_i$, at the region $0<q_\parallel<1$. 
For $q_\parallel\ge 1$, on the other hand, $z_r$ decreases with the increase
of $\kappa_i$. As for the imaginary part, $z_i$, the left panel of Fig.
\ref{fig7} shows that the most noticeable feature is that the maximum $z_i$
increases from $z_i\simeq 0.22$ to $z_i\simeq 0.28$ when $\kappa_i$ is
changed changed between 2.5 and 30.0. That is, the growth rate of the 
ion-cyclotron instability is increased with the decrease of the energetic
tail in the ion distribution.

\begin{widetext}

\begin{figure}[H]
\includegraphics{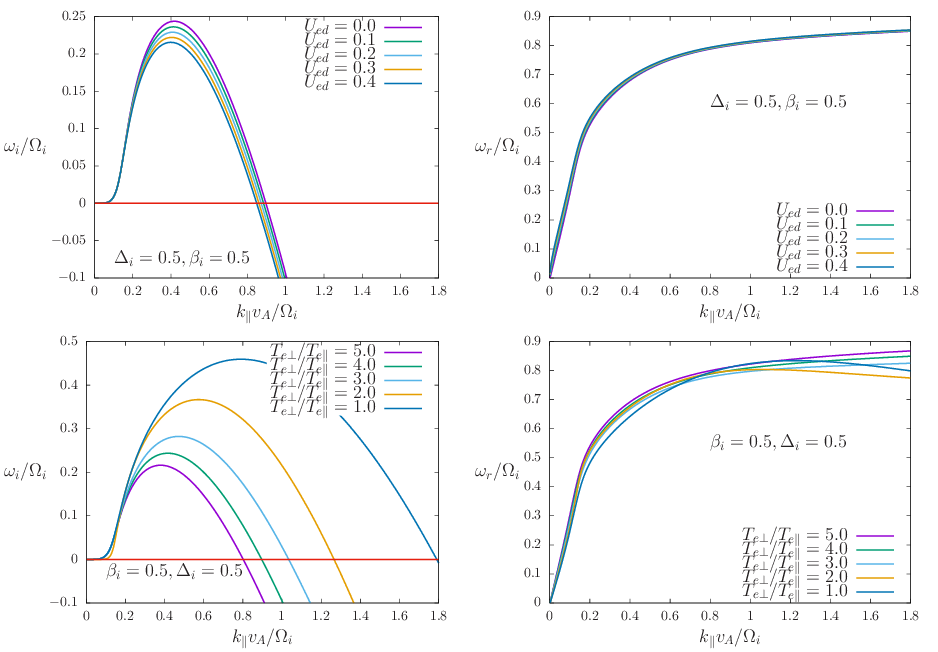}
\caption{
The left panels show the imaginary part of the normalized frequency of unstable
parallel propagating electromagnetic waves, $z_i$, vs. normalized wave number 
$q_\parallel$. 
The panels to the right show the corresponding real part of the normalized 
frequency, $z_r$, vs. normalized wave number $q_\parallel$.
Ions with a subtracted-Kappa distribution (SK), with $\kappa_i=3.0$, 
$T_{i\perp}/T_{i\parallel}= 4.0$, $\beta_i=0.5$, and $\Delta_i=0.5$;
Top panels: Electrons with a drifting BK distribution, with $\kappa_e=3.0$ and
$T_{e\perp}/T_{e\parallel}= 4.0$, for several values of the normalized electron 
drift velocity, $U_{ed}$. 
Bottom panels: Electrons with a BK distribution ($U_{ed}=0$), with 
$\kappa_e=3.0$, and several values of the temperature ratio
$T_{e\perp}/T_{e\parallel}$.
}
\label{fig8}
\end{figure}
\end{widetext}

The final part of this numerical investigation on the effect of plasma
parameters on the ion-cyclotron instability is the investigation of changes
in the electron distribution function. In the top panels of Fig. \ref{fig8}
one can find some results which show the influence of a drift velocity on the
electron distribution function. 
The ions are described by a subtracted-Kappa distribution (SK), with 
$\kappa_i=3.0$, 
$T_{i\perp}/T_{i\parallel}= 4.0$, $\beta_i=0.5$, and $\Delta_i=0.5$, and the
electrons by drifting BK distribution, with $\kappa_e=3.0$ and
$T_{e\perp}/T_{e\parallel}= 4.0$, with several values of $U_{de}$. 
It is seen in the top panels of Fig. \ref{fig8} that the effect of $U_{de}$
on the value of $z_r$ can be considered negligible, and that the effect on
the value of $z_i$ is a small reduction in the maximum growth rate and on the
range of the unstable wave numbers.
On the other hand, the effect of the thermal anisotropy in the electron 
distribution is much more significant, as shown in the bottom panels of 
Fig. \ref{fig8}. For instance, the left panel shows that the unstable range
is $0.1<q_\parallel<0.9$, with maximum growth rate $z_i\simeq 0.23$, when the
electron anisotropy is the same as the ion anisotropy. The range of unstable
wave numbers grows when the electron anisotropy decreases, and when the 
electron distribution becomes isotropic ($T_{e\perp}/T_{e\parallel}=1.0$) the
upper limit of the unstable range is $q_\parallel\simeq 1.8$, with maximum
growth rate $z_i\simeq 0.46$. These result show that, for ions described by
SK distributions, the effect of thermal anisotropy in the electron 
distribution is to decrease the magnitude and wave number range of the 
ion-cyclotron instability, effect similar to the effect observed in the
case of ions with BK distributions, without the loss-cone feature
\cite{pl:ZiebellG17}.
  
\section{Final remarks}
\label{sec:final}

This paper presents results obtained with numerical solution of the
dispersion relation for electromagnetic waves propagating parallel to the
direction of an ambient magnetic field. We have considered plasma parameters
which are representative of space plasma environments, and have solved the 
dispersion relation considering the ion population described by a mathematical
form of distribution which has recently appeared in the literature, the
subtracted-Kappa distribution \cite{SummersS2025}. 

The analysis has been made considering two different conditions regarding the
thermal anisotropy of the ion distribution, namely
$T_{i\perp}/T_{i\parallel}<1$ and 
$T_{i\perp}/T_{i\parallel}>1$. 
As it is known, for 
$T_{i\perp}/T_{i\parallel}<1$ the firehose instability may occur. 
We have chosen the value $T_{i\perp}/T_{i\parallel}=0.4$, and the results 
obtained have shown that for the parameters $\Delta_i=0$ and $\beta_i=0$, 
case in which the ion distribution is a BK distribution, the
instability occurs for normalized wavenumber between 0 and 1.5, with maximum 
growth rate $z_i\simeq 0.12$ at $q\simeq 0.4$. For increasing values of the 
parameter $\beta_i$, when the distribution gradually changes to a loss-cone
distribution, the instability is greatly reduced, being confined to the region
$0.1<q<0.5$ for $\beta_i=0.8$, with maximum growth rate $z_i\simeq 0.02$ at
$q\simeq 0.3$. For increasing values of $\Delta_i$, the effect of $\beta_i$ 
on the instability growth rate becomes less prominent.

We have also discussed the effect of the parameter $\Delta_i$ on the FH instability,
for different values of $\beta_i$. We have seen that the increase of $\Delta_i$
leads to increase in the FH instability, but the effect depends on the value of
$\beta_i$. 
For small value of $\beta_i$, the effect of $\Delta_i$ is very small, but it
increases for increasing values of $\beta_i$. 
As example, we have seen that for $\beta_i=0.9$, the increase of
$\Delta_i$ from 0.0 to 0.8 leads to increase in the maximum value of $z_i$, 
from $z_i \simeq 0.01$ to $z_i\simeq 0.1$, with the upper range of the 
unstable region changing from $q\simeq 0.4$ to $q\simeq 1.2$.
  
When investigating the effect of the parameter $\kappa_i$, controlling the 
occurrence of an extended power-law tail in the ion distribution, we have seen
that the increase of $\kappa_i$ leads to increase in the maximum growth rate
of the FH instability, 
while maintaining the range of unstable wave numbers.

Another feature investigated regarding the firehose instability has been the
effect of characteristics of the electron distribution on the magnitude and
range of the instability. We have seen that, for ions with a SK distribution
the presence of a drift velocity in the electron distribution contributes
to decrease of the FH instability, both in magnitude and in range of unstable
wave numbers. Moreover, the increase of the thermal anisotropy of the electron
population contributes to increase of the FH instability, both in range and
in the magnitude of the growth rate.

Other situation considered in the paper has been the occurrence of ion thermal
anisotropy such that $T_{i\perp}/T_{i\parallel}>1$, condition in which the 
ion-cyclotron instability may occur. 
We have chosen the value $T_{i\perp}/T_{i\parallel}=4.0$. The results 
obtained have shown that in the case in which the ion distribution is a BK 
distribution, with $\Delta_i=0$ and $\beta_i=0$, the IC instability occurs for 
normalized wavenumber between 0.1 and 0.75, with maximum 
growth rate $z_i\simeq 0.16$ at $q\simeq 0.438$. For increasing values of the 
parameter $\beta_i$, when the distribution gradually changes to a loss-cone
distribution, the instability greatly enhances, enlarging the unstable region
to $0.1<q<1.18$ for $\beta_i=0.8$, with maximum growth rate $z_i\simeq 0.39$ at
$q\simeq 0.5$. The effect of $\beta_i$ on the IC instability has been seen to
be opposite to the effect seen on the FH instability. However, as in the
case of FH instability, for increasing values of $\Delta_i$, the effect 
of $\beta_i$ on the growth rates of the IC instability becomes less preminent.

Regarding the effect of the parameter $\Delta_i$ on the IC instability,
for fixed values of $\beta_i$, the results obtained show that the increase of 
$\Delta_i$ leads to decrease in the IC instability, in magnitude of the
growth rate and in range of unstable wavenumbers. In this sense the effect
of $\Delta_i$ is contrary to that seen in the case of FH instabilities.
However, as for the FH instability, on the IC instability the magnitude of 
the effect of $\Delta_i$ is small for small values of $\beta_i$, and increases 
when $\beta_i$ is increased. 

Regarding the effect of existence of an extended power-law tail in the ion 
distribution, described by the parameter $\kappa_i$, we have seen 
that the increase of $\kappa_i$, i.e, reduction of the extended tail, 
leads to increase in the maximum growth rate of the IC instability. As in the
case of the FH instability, in the case of IC instability the increase
in $\kappa_i$ does not affect the upper limit of the range of unstable wave
numbers. However, in the case of IC, the upper limit of the unstable range is
slightly reduced, so that the range of unstable wave numbers decreases when
the extended tail in the ion distribution is decreased.

Regarding the effect of characteristics of the electron distribution on the 
magnitude and range of the IC instability, we have seen that in the case of
ions with a SKd distribution the increase in the drift velocity of the 
contributes to decrease of the growth rates of the IC instability. However, the
magnitude of the effect in the case of IC instability is less significant
than in the case of FH instability. Our results show a reduction of nearly 
$20\%$ in the maximum value of $z_i$ for the IC instability, when the 
normalized electron drift velocity changes from $U_{de}=0.0$ to $U_{de}=0.4$,
while the same change in drift velocity lead to a reduction by nearly a 
factor or 10 in the case of the FH instability. 
Finally, while for the FH instability the increase of the thermal anisotropy 
of the electron population contributes to greatly increase the
instability, both in range and in the magnitude of the growth rate, in the case
of the IC instability the effect is inverse. The increase of 
$T_{e\perp}/T_{e\parallel}$ contributes to decrease the magnitude of the growth
rate of the IC instability, as well as to decrease the range of unstable 
wavenumbers.

Future extensions of the work presented in this paper may take into account the
possibility of wave propagation at direction oblique to the ambient magnetic.
We are investigating the subject, and intend to present our findings in a
forthcoming publication.

\begin{acknowledgments}
DS acknowledges support from a Discovery
Grant of the Natural Sciences and Engineering Research Council of
Canada (Grant No. 2016-04372).
LFZ acknowledges support from CNPq (Brazil), grant No. 303189/2022-3.
This study was financed in part by the Coordena\c{c}\~ao de Aperfei\c{c}oamento
de Pessoal de N\'{\i}vel Superior - Brasil (CAPES) - Finance Code 001.
\end{acknowledgments}

\begin{itemize}
\item ORCID:\\
Danny Summers: 0000-0002-9104-242X \\
Luiz F. Ziebell: 0000-0003-0279-0280 \\
Rudi Gaelzer: 0000-0001-5851-7959
\end{itemize}

\section*{Declarations}

\begin{itemize}

\item Conflict of interest

The authors have no financial or non-financial interests to declare.

\item Data availability

The data that support the findings of this study are available within the
article.

\end{itemize}

\appendix

\section{Evaluation of the quantity $J(n,m,h,f_{a0})$}

Equation (\ref{disprel}) shows 
that the dispersion relation requires the quantity $J(n,m,h,f_{a0})$
evaluated for $n=s$, $m=1$ and $h=0$. 
Using the operator $L$ defined in Eq. \eqref{J,def} 
applied to the auxiliary distribution given by Eq.
(\ref{fa,BK}), we obtain
\begin{eqnarray}
&&L(f_{a}^{(BK)}(u_\parallel,u_\perp; u_{a\kappa\parallel},u_{a\kappa\perp}))\\
&&=\left(-\frac{2u_\perp}{\gamma}\frac{\kappa_a+\alpha}{\kappa_a}\right)
\left[\frac{\gamma}{u_{a\kappa\perp}^2}
-\frac{q_\parallel}{z}\left(\frac{u_\parallel}{u_{a\kappa\perp}^2}
-\frac{u_\parallel-u_{da}}{u_{a\kappa\parallel}^2}\right)
\right]\nonumber\\
&&\times\left(1
+\frac{(u_\parallel-u_{da})^2}{\kappa_a u_{a\kappa\parallel}^2}
+\frac{u_\perp^2}{\kappa_a u_{a\kappa\perp}^2}\right)^{-1}
f_{a}^{(BK)}
\nonumber
\end{eqnarray}

Using this expression taking into account the non relativistic approximation 
($\gamma \simeq 1$), the required $J$ expression becomes the following,
\begin{eqnarray}
&&J(s,1,0;f_{a}^{(BK)}
(u_\parallel,u_\perp;u_{a\kappa\parallel},u_{a\kappa\perp}))\\
&&= - 2z \frac{(2\pi)n_{a 0}}
{\pi^{3/2} \kappa_a^{3/2}
u_{a\kappa\perp}^2 u_{a\kappa\parallel}}\frac{\kappa_a+\alpha}{\kappa_a}
\frac{\Gamma(\kappa_a+\alpha)}
{\Gamma(\kappa_a+\alpha-3/2)}\nonumber\\
&&\times\int_{-\infty}^\infty du_\parallel\,
\frac{1}{z - s \Omega_a/\Omega_i - q_\parallel u_\parallel}
\nonumber \\
&&\times
\int_0^\infty du_\perp \,u_\perp^{3}
\left[\frac{1}{u_{a\kappa\perp}^2}
-\frac{q_\parallel}{z}
\left(\frac{u_\parallel}{u_{a\kappa\perp}^2}
-\frac{u_\parallel-u_{da}}{u_{a\kappa\parallel}^2}\right)
\right]\nonumber\\
&&\times\left(1 +\frac{(u_\parallel-u_{da})^2}{\kappa_a u_{a\kappa\parallel}^2}
+\frac{u_\perp^2}{\kappa_a u_{a\kappa\perp}^2}
\right)^{-(\kappa_a+\alpha+1)} \nonumber.
\end{eqnarray}

Proceeding, we define $\displaystyle 
u_{\parallel,res}=\frac{z - s \Omega_a/\Omega_i}{q_\parallel}$,
$\displaystyle \eta_a= 1+\frac{(u_\parallel-u_{da})^2}{\kappa_a 
u_{a\kappa\parallel}^2}$ and $\displaystyle t=\frac{u_\perp^2}{\kappa_a
\eta_a u_{a\kappa\perp}^2}$, and obtain
\begin{eqnarray}
&&J(s,1,0;f_{a}^{(BK)}
(u_\parallel,u_\perp;u_{a\kappa\parallel},u_{a\kappa\perp}))\\
&&=\frac{2\omega}{v_\ast k_\parallel} \frac{(2\pi)n_{a 0}}
{\pi^{3/2} \kappa_a^{3/2} 
u_{a\kappa\perp}^2 u_{a\kappa\parallel}}
\frac{\kappa_a+\alpha}{\kappa_a}
\frac{\Gamma(\kappa_a+\alpha)}
{\Gamma(\kappa_a+\alpha-3/2)}\nonumber\\
&&\times\int_{-\infty}^\infty du_\parallel\, 
\frac{1}{u_\parallel-u_{\parallel,res}}
\left(1+\frac{(u_\parallel-u_b)^2}
{\kappa_a u_{a\kappa\parallel}^2}\right)^{-(\kappa_a+\alpha+1)}
\nonumber\\
&&\times
\frac{(\kappa_a \eta_a u_{a\kappa\perp}^2)^2}{2}
\left[\frac{1}{u_{a\kappa\perp}^2}
-\frac{q_\parallel}{z}\left(\frac{u_\parallel}{u_{a\kappa\perp}^2}
-\frac{u_\parallel-u_{da}}{u_{a\kappa\parallel}^2}\right)
\right]\nonumber\\
&&\times
\int_0^\infty dt \,t
\left(1+t\right)^{-(\kappa_a+\alpha+1)}\nonumber
\end{eqnarray}

The integral along the $t$ variable can be performed using the following
expression,
\begin{equation}
\label{integral-t}
\int_0^\infty dt~\frac{t^{z-1}}{(1+t)^{w+z}}
=\frac{\Gamma(z)\Gamma(w)}{\Gamma(w+z)}, 
\quad (\Re z>0, \Re w>0),
\end{equation}
and therefore
\begin{eqnarray}
&&J(s,1,0;f_{a}^{(BK)}
(u_\parallel,u_\perp;u_{a\kappa\parallel},u_{a\kappa\perp}))\\
&&=\frac{2n_{a 0}}
{\pi^{1/2} \kappa_a^{1/2}}
\frac{\Gamma(\kappa_a+\alpha-1)}{\Gamma(\kappa_a+\alpha-3/2)}
\nonumber\\
&&\times
\int_{-\infty}^\infty ds\,
\frac{1}{s-\hat{\zeta}_a^s}
\left(1+\frac{s^2}{\kappa_a}\right)^{-(\kappa_a+\alpha-1)}\nonumber\\
&&\times \left[\zeta_a^0
-s\left(1-\frac{u_{a\kappa\perp}^2}{u_{a\kappa\parallel}^2}\right)
-\frac{u_{da}}{u_{a\kappa\parallel}}
\right]\nonumber
\end{eqnarray}
where
\begin{eqnarray*}
s=\frac{u_\parallel-u_{da}}{u_{a\kappa\parallel}},\quad
\hat{\zeta}_a^n=\frac{z - n r_a -q_\parallel u_{da}}
{q_\parallel u_{a\kappa\parallel}},\quad
\zeta_a^0= \frac {z}{q_\parallel u_{a\kappa\parallel}}.
\end{eqnarray*}

Proceeding, we write
\begin{eqnarray}
&&J(s,1,0;f_{a}^{(BK)}
(u_\parallel,u_\perp;u_{a\kappa\parallel},u_{a\kappa\perp}))\\
&&= \frac{(2\pi)n_{a 0}}
{\pi^{3/2} \kappa_a^{1/2}}
\frac{\Gamma(\kappa_a+\alpha-1)}{\Gamma(\kappa_a+\alpha-3/2)}
\nonumber\\
&&\times \Biggl\{
-2\int_{0}^\infty ds\,
\left(1+\frac{s^2}{\kappa_a}\right)^{-(\kappa_a+\alpha-1)}
\left(1-\frac{u_{a\kappa\perp}^2}{u_{a\kappa\parallel}^2}\right)
\nonumber\\
&& + \int_{-\infty}^\infty ds\,
\frac{1}{s-\hat{\zeta}_a^n}
\left(1+\frac{s^2}{\kappa_a}\right)^{-(\kappa_a+\alpha-1)}\nonumber\\
&&\times\left[\zeta_a^0
-\hat{\zeta}_a^n
\left(1-\frac{u_{a\kappa\perp}^2}{u_{a\kappa\parallel}^2}\right)
-\frac{u_{da}}{u_{a\kappa\parallel}}
\right] \Biggr\},\nonumber
\end{eqnarray}

For the integral in the first term between the curly brackets, we change 
variable using $t= s^2/\kappa_a$, and therefore obtain
\begin{eqnarray}
\label{JBKd,h=0,1}
&&J(s,1,0;f_{a}^{(BK)}
(u_\parallel,u_\perp;u_{a\kappa\parallel},u_{a\kappa\perp}))\\
&&= \frac{(2\pi)n_{a 0}}
{\pi^{3/2} \kappa_a^{1/2}}
\frac{\Gamma(\kappa_a+\alpha-1)}{\Gamma(\kappa_a+\alpha-3/2)}
\Biggl\{ -\kappa_a^{1/2}
\left(1-\frac{u_{a\kappa\perp}^2}{u_{a\kappa\parallel}^2}\right)
\nonumber\\
&&\times 
\int_{0}^\infty dt\,\frac{1}{t^{1/2}}
\left(1+t\right)^{-(\kappa_a+\alpha-1)}
\nonumber\\
&& + \pi^{1/2}\kappa_a^{1/2}
\frac{\Gamma(\kappa_a-1/2)}{\Gamma(\kappa_a)}
Z_{\kappa_a}^{(\alpha-1)}(\hat{\zeta}_a^n)\nonumber\\
&&\times\left[\zeta_a^0
-\hat{\zeta}_a^s
\left(1-\frac{u_{a\kappa\perp}^2}{u_{a\kappa\parallel}^2}\right)
-\frac{u_{da}}{u_{a\kappa\parallel}}
\right] \Biggr\},\nonumber
\end{eqnarray}
where
we have utilized the definition the plasma dispersion function for 
$\kappa$ distributions, of order $m$,
\begin{eqnarray}
&&Z_\kappa^{(m)}(\xi)= \frac{1}{\pi^{1/2}}\frac{\Gamma(\kappa)}{\kappa^{1/2}
\Gamma(\kappa-1/2)}\\
&&\times\int_{-\infty}^\infty
\frac{ds}{(s-\xi)(1+s^2/\kappa)^{\kappa+m}},\nonumber
\label{Zkappam}
\end{eqnarray}

This dispersion function
reduces to the distribution defined by Summers and Thorne
\cite{SummersT91} in the case $m=1$. It is particularly
useful for numerical applications 
that it can be written in terms of the Gauss hypergeometric function 
$_2F_1(a,b,c,z)$, as follows,
\begin{eqnarray}
&&Z_\kappa^{(m)}(\xi)= \frac{i\Gamma(\kappa)\Gamma(\kappa+m+1/2)}{\kappa^{1/2}
\Gamma(\kappa-1/2)\Gamma(\kappa+m+1)}\\
&&\times {_2F_1}\left[1,2\kappa+2m;\kappa+m+1;
\frac{1}{2}\left(1+\frac{i\xi}{\kappa^{1/2}}\right)\right],\nonumber
\end{eqnarray}
for $\kappa>-m-1/2$.

The integral in the first term of Eq. (\ref{JBKd,h=0,1}) can be evaluated
with use of Eq. (\ref{integral-t}), Performing the integration, and simplifying
the ensuing expression, we obtain
\begin{eqnarray}
\label{JBKd,h=0,2}
&&J(s,1,0;f_{a}^{(BK)}
(u_\parallel,u_\perp;u_{a\kappa\parallel},u_{a\kappa\perp}))\\
&&= 2n_{a 0}
\frac{\Gamma(\kappa_a+\alpha-1)}{\Gamma(\kappa_a+\alpha-3/2)}
\nonumber\\
&&\times \Biggl\{
-\frac{\Gamma(\kappa_a+\alpha-3/2)}
{\Gamma(\kappa_a+\alpha-1)}
\left(1-\frac{u_{a\kappa\perp}^2}{u_{a\kappa\parallel}^2}\right)\nonumber\\
&& + \frac{\Gamma(\kappa_a-1/2)}{\Gamma(\kappa_a)}
\left[\zeta_a^0
-\hat{\zeta}_a^s
\left(1-\frac{u_{a\kappa\perp}^2}{u_{a\kappa\parallel}^2}\right)
-\frac{u_{da}}{u_{a\kappa\parallel}}
\right]
Z_{\kappa_a}^{(\alpha-1)}(\hat{\zeta}_a^s)
\Biggr\},\nonumber
\end{eqnarray}


\end{document}